\documentclass[a4paper, anonymous]{article}

\usepackage{ISCSLP2022}
\usepackage{multirow}
\usepackage{booktabs}
\usepackage{array}
\usepackage{float}
\usepackage{ulem}
\usepackage{color}
\newcommand{\tabincell}[2]{\begin{tabular}{@{}#1@{}}#2\end{tabular}}

\title{Low Pass Filtering and Bandwidth Extension for Robust Anti-spoofing Countermeasure Against Codec Variabilities}

\name{Yikang Wang$^{1,2}$, Xingming Wang$^{2,3}$, Hiromitsu Nishizaki$^{1}$, Ming Li$^{2,3*}$\thanks{$^*$Corresponding author.\\This research is funded in part by the National Natural Science Foundation of China (62171207). Many thanks for the computational resource provided by the Advanced Computing East China Sub-Center.}}
\address{
  $^1$Integrated Graduate School of Medicine, Engineering, and Agricultural Sciences, \\University of Yamanashi, Japan\\
  $^2$Data Science Research Center, Duke Kunshan University, China\\
  $^3$School of Computer Science, Wuhan University, China}
\email{ming.li369@duke.edu\vspace{-1.6em}}

\begin{document}
\maketitle

\begin{abstract}
A reliable voice anti-spoofing countermeasure system needs to robustly protect automatic speaker verification (ASV) systems in various kinds of spoofing scenarios. However, the performance of countermeasure systems could be degraded by channel effects and codecs. In this paper, we show that using the low-frequency subbands of signals as input can mitigate the negative impact introduced by codecs on the countermeasure systems. To validate this, two types of low-pass filters with different cut-off frequencies are applied to countermeasure systems, and the equal error rate (EER) is reduced by up to 25\% relatively. In addition, we propose a deep learning based bandwidth extension approach to further improve the detection accuracy. 
Recent studies show that the error rate of countermeasure systems increase dramatically when the silence part is removed by Voice Activity Detection (VAD), our experimental results show that the filtering and bandwidth extension approaches are also effective under the codec condition when VAD is applied.

\end{abstract}
\vspace{-0.6em}
\section{Introduction}

Automatic speaker verification (ASV) \cite{reynolds2000speaker,desplanques20_interspeech} has achieved good performance and is widely used in real life. As a biometric method, however, ASV systems are vulnerable against various spoofing attacks, e.g. speech synthesis, voice conversion, record and playback, etc. \cite{kamble2020advances,evans2014speaker}. 
Anti-spoofing countermeasure systems contribute to enhance the reliability of ASV systems by determining whether the input signal is genuine or spoofed.

Since 2015, the ASVspoof community has initiated and organized four consecutive biennial challenges to support the development of anti-spoofing countermeasure methods for ASV systems \cite{asvspoof15,asvspoof17,asvspoof19,asvspoof21}. Through this series of challenges, the ASV anti-spoofing field has established two main anti-spoofing countermeasure research scenarios and databases: logical access (LA) considers spoofing attacks from text-to-speech synthesis (TTS) and voice conversion (VC), physical access (PA) refers to attacks produced by recording replay \cite{tom18_interspeech,baumann2021voice}. 
In this paper, we focus on the LA scenario.
A typical countermeasure system consists of a front-end feature extractor and a back-end spoofing classifier. For the LA task, the most intuitive solution is to find the artifact traces existing in the speech from TTS and VC through the front-end operations, which can be used as a marker or cue to distinguish genuine speech from artifacts and help train the back-end detection network better.
Previous works show that artifacts of synthetic speech exist in multiple specific subbands \cite{sriskandaraja16_interspeech, yang2019significance}. More and more studies focus on the impact of 
subbands in countermeasure systems \cite{Zhang}.

The current strategies for using frequency subbands in countermeasure systems can be broadly classified into two categories. One uses transformations, trimming, fusion or other methods in the front-end feature extractor to transform the features to a specific domain, which emphasize the information in target subbands \cite{sriskandaraja16_interspeech, Zhang, tak20_interspeech, chettri20_odyssey}. The other one directly adopt the spectrogram as the features, use SpecAugment \cite{park19e_interspeech} or similar data augmentation techniques to randomly or specifically block a portion of the frequency bands or time frames to reduce the overfitting of the back-end classifier \cite{lavrentyeva19_interspeech, beike_add2022}.

Zhang et al. point out that the high-frequency part of the spectrogram may lead to overfitting of the back-end neural network, increasing it's risk of making incorrect judgments in the face of unknown spoofing attacks \cite{Zhang}. 
Moreover, Tomilov et al. find that a data augmentation such as feeding training data into a filter to emulate the magnitude responses of codecs can yield better anti-spoofing countermeasure performance \cite{tomilov21_asvspoof}. 

However, it's not clear that whether low pass filtering is always useful with different bandwidths or scenarios, and whether the re-estimated high frequency information from a deep learning based up-sampling approach could bring some additional gain. This paper investigates the optimal cutoff frequency in terms of low pass filters against the codec variabilities.
Moreover, inspired by Tomilov et al's work \cite{tomilov21_asvspoof}, we use a conventional low-pass filter to obtain a subband signal instead of trimming the spectrogram directly. In addition, we further demonstrate the gain when using a deep learning based bandwidth extension technique to restore the wide band signal from the low pass filtered narrow band speech.

As shown in Figure~\ref{fig:system_archi}, we investigate the performance of countermeasure systems based on four front-end feature extraction methods and two convolutional neural networks (CNN) back-end classifiers. We also explore the performance of our countermeasure systems after applying a Voice Activity Detection (VAD) module. 
Our contributions are mainly threefold:
\begin{figure*}[th]
  \centering
  \includegraphics[width=0.98\linewidth]{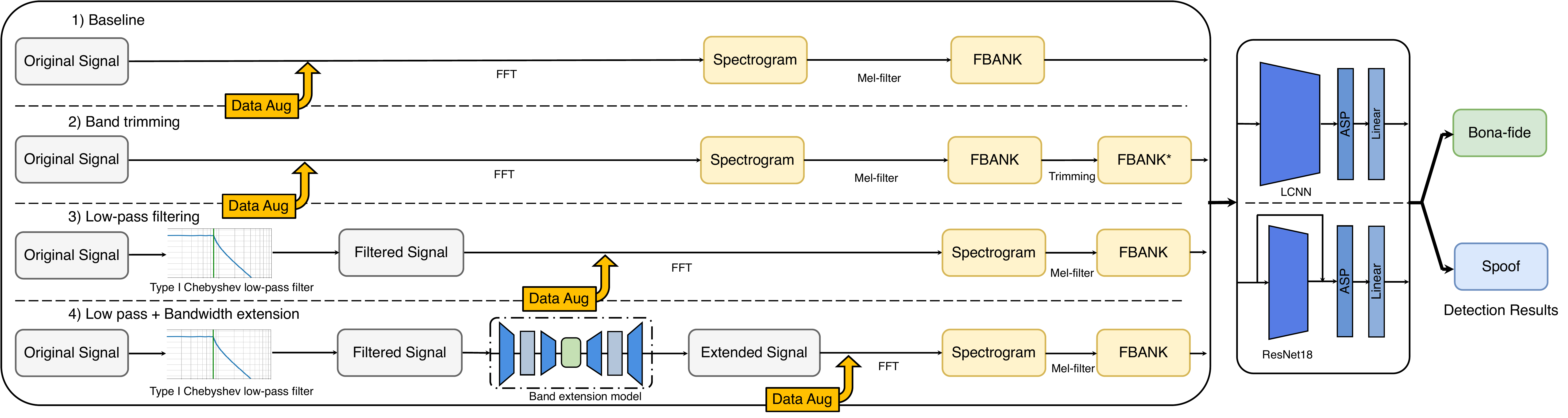}
  \caption{The overview of the proposed front-end signal processing methods.\vspace{-1.6em}}
  \label{fig:system_archi}
\end{figure*}
\begin{itemize}

\item We investigate the differences when using low-frequency subbands at the system's input features in two different databases and find out that high frequency subbands are more vulnerable against the codecs.
\item We use extensive experimental results to validate the optimal cutoff frequency for our countermeasure systems.
\item We first utilize a deep learning based bandwidth extension technique on the down-sampled signal in the countermeasure system, and suggest that the additional bandwidth extension module can be effective on the valid speech part when a VAD is applied.
\end{itemize}

\section{Database and Methodology}

This section describes the database and illustrates the detail of our front-end signal processing methods shown in Figure~\ref{fig:system_archi}.
\vspace{-0.7em}
\subsection{Databases}
The experiments in this work are mainly conducted on the ASVspoof2019 LA database \cite{asvspoof19} and the ASVspoof2021 LA database \cite{asvspoof21}. Both databases are based upon the Voice Cloning Toolkit (VCTK) corpus \cite{vctk}. 
The ASVspoof2019 LA database was created using utterances from 107 speakers (46 male, 61 female). The set of 107 speakers is partitioned into three speaker-disjoint sets for training, development, and evaluation. The spoofed utterances were generated using four TTS and two VC algorithms in the training and development sets, while 13 TTS/VC algorithms are used in the evaluation subset, 11 of which are unknown for training and development. The ASVspoof2021 LA database remains the training and development data unchanged and only proposes a new evaluation subset that contains attacks using the same simulation methods as the ASVspoof2019 LA evaluation subset. The ASVspoof2021 LA evaluation subset consists of the data in various telephone transmission systems, including Voice over Internet Protocol (VoIP) and the Public Switched Telephone Networks (PSTN), thus exhibiting real-world signal transmission channel effects. These effects could generate artifacts different from spoofing data that affect the accuracy of the countermeasure system. These conditions make the ASVspoof2021 LA evaluation subset an excellent platform for evaluating countermeasure systems' generalization capability and robustness against channel effects and codec variabilities. The contents of the two databases are summarized in Table~\ref{tab:database}.

\begin{table}[tb]
  \caption{Number of utterances in ASVspoof 2019 and 2021 LA database.\vspace{-0.6em}}
  \label{tab:database}
 \centering
\setlength{\tabcolsep}{1.25mm}{\begin{tabular}{lllll} 
\toprule
            & \multicolumn{2}{c}{\textbf{19LA }} & \multicolumn{2}{c}{\textbf{21LA }}  \\ 
\cmidrule(lr){2-3}\cmidrule(lr){4-5}
Subset      & Bona-fide & Spoof                  & Bona-fide & Spoof                   \\ 
\hline
Training    & 2,580     & 22,800                 & -         & -                       \\
Development & 2,548     & 22,296                 & -         & -                       \\
Evaluation  & 7,355    & 63,882                & 14,816    & 166,750                 \\
\bottomrule
\end{tabular}}
\vspace{-1.6em}
\end{table}

\vspace{-0.7em}
\subsection{Baseline}

We adopt the log Mel-filter bank energy (FBANK) as the acoustic feature in all our experiments. The Fast Fourier Transform (FFT) spectrogram is extracted with 1024 window length and 128 hop length while the Blackman window is used. Then we set the number of Mel-filters to 80 dimensions. Due to the different speech lengths, each spectrogram's length is truncated or concatenated into 3 to 5 seconds. Finally, the $80*frames$-dimensional FBANK features are obtained.

For the backend classifier, we investigated two main CNN networks, ResNet18 \cite{he2016deep} and LCNN \cite{lavrentyeva19_interspeech}. Then we add the attentive statistics pooling layer (ASP) \cite{okabe18_interspeech, desplanques20_interspeech} at the end of the models to make the model more effective in capturing utterance-level acoustic feature changes. The ResNet18 and the LCNN are commonly and widely used systems in anti-spoofing tasks \cite{wang21_asvspoof}. We use softmax cross entropy as the loss function of the classifier. The architectures of the models are shown in Tables~\ref{tab:res18} and~\ref{tab:lcnn}, respectively.
\begin{table}[tb]
    \footnotesize
    \caption{ The architecture of ResNet18, $\mathbf{C}$ denotes the convolutional layer, $\mathbf{S}$ denotes the shortcut convolutional layer.}
    \label{tab:res18}
    \centering
    \begin{tabular}[c]{@{\ \ \ }l@{\ \ \ }c@{\ \ \ }l@{\ \ \ }}
        \toprule
        \textbf{Layer} & \textbf{Output Size} & \textbf{Structure(kernal size, stride)} \\
        \midrule
        Conv1 & $16 \times D \times L$ & $\mathbf{C}(3\times 3, 1)$ \\
        \midrule
        \tabincell{l}{Residual\\Layer 1} & $16 \times D \times L$ & $\begin{bmatrix}
            \mathbf{C}(3\times 3, 1) \\
            \mathbf{C}(3\times 3, 1)
        \end{bmatrix} \times 2$ \\
        \midrule
        \tabincell{l}{Residual\\Layer 2} & $32 \times \frac{D}{2} \times \frac{L}{2}$ & $\begin{bmatrix}
            \mathbf{C}(3\times 3, 2) \\
            \mathbf{C}(3\times 3, 1) \\
            \mathbf{S}(1\times 1, 2)
        \end{bmatrix} \begin{bmatrix}
            \mathbf{C}(3\times 3, 1) \\
            \mathbf{C}(3\times 3, 1)
        \end{bmatrix}\times 2$ \\
        \midrule
        \tabincell{l}{Residual\\Layer 3} & $64 \times \frac{D}{4} \times \frac{L}{4}$ & $\begin{bmatrix}
            \mathbf{C}(3\times 3, 2) \\
            \mathbf{C}(3\times 3, 1) \\
            \mathbf{S}(1\times 1, 2)
        \end{bmatrix} \begin{bmatrix}
            \mathbf{C}(3\times 3, 1) \\
            \mathbf{C}(3\times 3, 1)
        \end{bmatrix}\times 2$ \\
        \midrule
        \tabincell{l}{Residual\\Layer 4} & $128 \times \frac{D}{8} \times \frac{L}{8}$ & $\begin{bmatrix}
            \mathbf{C}(3\times 3, 2) \\
            \mathbf{C}(3\times 3, 1) \\
            \mathbf{S}(1\times 1, 2)
        \end{bmatrix} \begin{bmatrix}
            \mathbf{C}(3\times 3, 1) \\
            \mathbf{C}(3\times 3, 1)
        \end{bmatrix}\times 2$ \\
        \midrule
        Pooling & $128 \times \frac{D}{8}$ & Attentive Statistics Pooling \\
        Linear & $128$ & Fully Connected Layer\\
        Linear & $2$ & Fully Connected Layer\\
        \bottomrule
    \end{tabular}
\vspace{-1em}
\end{table}
\begin{table}[tb]
    \centering
    \caption{The architecture of LCNN.\vspace{-0.6em}}
    \label{tab:lcnn}
    \setlength{\tabcolsep}{-0.2mm}{\begin{tabular}{lcl} 
    \toprule
    \textbf{Layer} & \textbf{Structure(kernal size, stride)}     & \textbf{Output Size}                        \\ 
    \midrule
    &Same as the first 28 layers of LCNN \cite{lavrentyeva19_interspeech}& \\
    \midrule
    Pooling & BiLSTM & $80+80$\\
    Pooling & Attentive Statistics Pooling & $320$ \\
    Linear  &Fully Connected Layer& $128$ \\
    Linear  & Fully Connected Layer& $2$\\
    \bottomrule
    \end{tabular}}
\vspace{-1em}
\end{table}

\subsection{Band trimming}

Band trimming means cropping a particular dimension of the Mel-filter bank from the complete FBANK to make it consistent with the spectrogram subbands. According to the Nyquist frequency characteristic, the spectrogram of the speech with a sampling rate of 16k covers a bandwidth of 0-8k Hz, and we need to select the low-frequency subband cover 0-$\mathbb{F}$ Hz for training and testing the countermeasure system, which $\mathbb{F}$ here denotes the subband frequencies corresponding to 20\%, 30\%, 40\%, 50\%, 60\%, 70\% Nyquist frequencies. The equation describing the correlation between the number of low- and full-frequency FBANK dimensions \cite{cai2021unified} is shown as follows
\begin{equation}
  \lfloor N_L \rfloor = \lfloor N_F \ast \frac{log(1+f_L/700)}{log(1+f_F /700)}\rfloor.
  \label{eq1}
\end{equation}
Where $\lfloor*\rfloor$ indicates rounding down an element $*$. If the operation of band trimming is considered to be low-pass filtering of the signal, then $f_L$ denotes the cutoff frequency of this filter and $f_F$ refers to the Nyquist frequency. $N_L$ and $N_F$ represent the number of filter banks for the low- and full-frequency FBANK feature, respectively. For instance, when we need to select the low-frequency subband up to 50\% of the Nyquist frequency, the corresponding value of $f_F$, $f_L$, sample frequency, and $N_F$ are $8,000$, $4,000$, $16,000$, and $80$ respectively. Therefore, according to Equation~\ref{eq1}, we get $N_L=60.45$. The feature dimension must be an integer, so we floor $N_L$ to be 60 and set $f_L$ to 3,933.55 Hz. In other words, the spectrogram information from 3,933.55 Hz to 4,000 Hz is dropped out. With respect to Equation~\ref{eq1}, 20\%, 30\%, 40\%, 50\%, 60\%, 70\% of Nyquist frequencies corresponds to the FBANK indices 37, 47, 54, 60, 65, and 69.

\vspace{-0.7em}
\subsection{Low-pass filtering}
Considering the response curve shape, the computational complexity, and the consistency of the filter in the bandwidth extension front-end, the Chebyshev type I filter is chosen as the low-pass filter in our experiments.
We set the order of filter to 8, the maximum ripple of the Chebyshev type I low-pass filter to 0.05, and the critical frequencies to $\mathbb{F}$ Hz, which $\mathbb{F}$ is the same as the one mentioned in the last subsection. 
\vspace{-0.7em}
\subsection{Bandwidth extension}
Bandwidth extension is also known as audio upsampling or audio super-resolution. It usually aims to enhance speech audibility and improve audio fidelity by generating a wideband (WB) signal from a narrowband (NB) signal. In order to use extension models to enhance the performance of the countermeasure system in this study, we investigate some bandwidth extension methods \cite{jax2003artificial,zhang21k_interspeech,lee21c_interspeech}. Among them, Viet-Anh Nguyen et al. proposed a Transformer-aided UNet (TUNet)\footnote{Source code: https://github.com/NXTProduct/TUNet.} by employing a low-complexity transformer encoder on the bottleneck of a lightweight UNet \cite{Tunet}. Their experimental results on the VCTK corpus show that the TUNet achieves state-of-the-art performance in intrusive and non-intrusive metrics.

The workflow of the bandwidth extension front-end is shown on the left below of Figure~\ref{fig:system_archi}, where a Chebyshev Type I lowpass filter is used to preprocess the original 16k Hz signal into a low pass filtered signal but still at a sampling rate of 16k Hz. 
After that, the filtered signal is used as the input of the TUNet extension model \cite{Tunet}, and the extended signal containing 0-8k Hz spectrogram can be restored after network inference. Finally, the output of the extension model is changed to FBANK features and as the input of the back-end classifier. 

\section{Experimental Setup}

\subsection{Data Augmentation}
As shown in Figure~\ref{fig:system_archi}, in order to improve the robustness of the countermeasure classifier, we implemented data augmentation to add noise before FFT in the front-end. Motivated by the data augmentation methods used in the ASV system with the VoxCeleb database \cite{chung20b_interspeech,nagrani17_interspeech,chung18b_interspeech}, reverberation and background noise are added randomly to two-thirds of the input data. The noise data are obtained from the MUSAN \cite{musan2015} and RIR \cite{ko2017study} databases.

\vspace{-0.7em}
\subsection{Metric and training strategy}
The evaluation was performed in terms of Equal Error Rate (EER) as a metric, which indicates that the proportion of false acceptances is equal to the proportion of false rejections.



\begin{table*}[t]
\caption{The EER\% of countermeasure systems, when choosing 0.5 Nyquist frequency as the cutoff frequency. \textbf{Bold number} indicates the best performing result among the countermeasure systems corresponding to each back-end network in the 19LA or 21LA database.}
\label{tab:example3}
\centering
\begin{tabular}{llllllllll} 
\toprule
\multirow{2}{*}{\textbf{back-ends}} & \multirow{2}{*}{\textbf{front-ends}} & \multicolumn{4}{c}{\textbf{ASVspoof2021 LA}}              & \multicolumn{4}{c}{\textbf{ASVspoof2019 LA}}      \\ 
\cmidrule(lr){3-6}\cmidrule(lr){7-10}
               & & seed1 & seed10 & seed100 & Average    & seed1 & seed10 & seed100 & Average     \\ 
\hline
\multirow{4}{*}{ResNet18} & baseline  & 3.23  & 3.35   & 3.38    & 3.32       & 1.76  & 1.48   &\textbf{1.15}    & 1.46  \\
                          & band trimming   & 2.65  & \textbf{2.32}   & 2.44    & 2.47       & 1.37  & 1.66   & 1.46     & 1.5  \\
                          & low-pass filtering   & 3.05  & 3.03   & 3.05    & 3.04       & 1.34  & 1.7  & 1.57     & 1.53  \\
                          & low pass + bandwidth extension  & 2.35  & 2.57   & 2.72    & 2.54       & 1.37  & 1.38   & 1.6     & 1.45  \\ 
\hline
\multirow{4}{*}{LCNN}     & baseline  & 3.95  & 4.37   & 3.38    & 4.26       & 2.36  & 2.36   &2.5    & 2.41  \\
                          & band trimming   & 3.23  & \textbf{2.72}   & 2.76    & 2.90       & 2.03  & 1.75   & \textbf{1.72}     & 1.83  \\
                          & low-pass filtering   & 3.51  & 2.84   & 3.07    & 3.14       & 2.31  & 1.89   & 2.14     & 2.11  \\
                          & low pass + bandwidth extension  & 2.94  & 2.97   & 3.01    & 2.91       & 2.05  & 2.05   & 1.84     & 1.98  \\
\bottomrule
\end{tabular}
\vspace{-1.3em}
\end{table*}
For training the countermeasure systems, Adam \cite{kingma2014adam} was used as the optimizer with $\beta_1=0.9$, $\beta_2=0.999$, $\epsilon=10^{-8}$, and weight decaying $10^{-4}$. For the CNN classifiers, the learning rate increase linearly for the first four warm-up epochs and then is initialized to 0.001 starting from the fifth epoch. The learning rate is scheduled as the PyTorch ReduceLROnPlateau function\footnote{https://pytorch.org/docs/stable/generated/torch.optim.lr\_sched\\uler.ReduceLROnPlateau.html}, reducing the learning rate when the metric has stopped improving for ten epochs. Each experiment uses one NVIDIA RTX A6000 GPU, and for efficiency, we set the batch to 400, with 100 epochs of training per model.
\begin{figure}[tb]
  \flushleft 
  \includegraphics[width=0.94\linewidth]{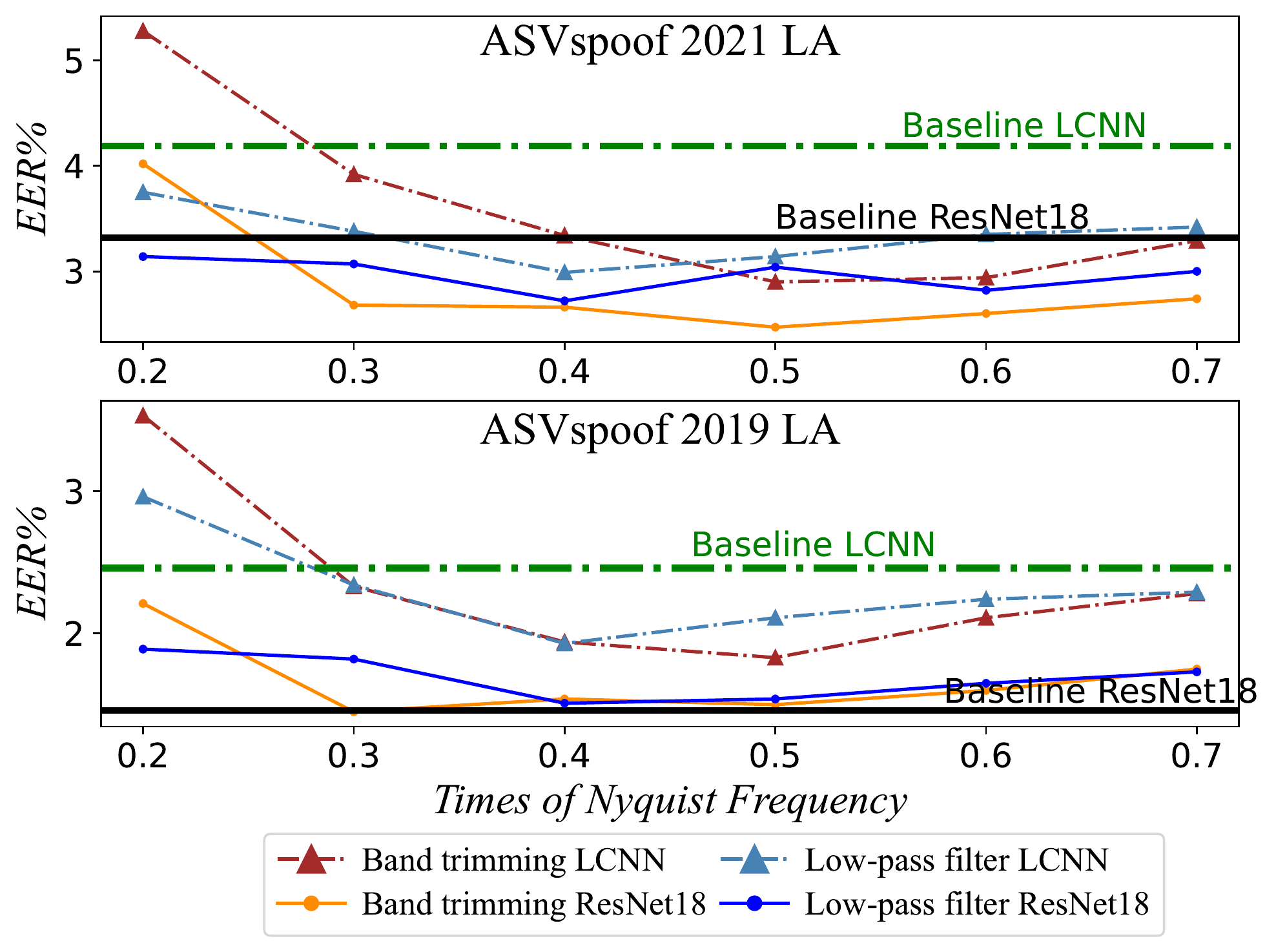}
  \caption{\vspace{-0.1em}Performance comparison of two subband front-ends (band trimming and low-pass filtering) with the baseline front-end countermeasure system at different cut-off frequencies.\vspace{-1.6em}}
  \label{fig:result1}
\end{figure}
\section{Experimental Results and Discussion}
\subsection{Countermeasure systems with different front-end}

From Figure~\ref{fig:result1}, it seems a cutoff frequency of 0.5 is a good candidate for comparison experiments. The experimental results of all eight countermeasure systems are shown in Table~\ref{tab:example3}, in which the cutoff frequencies were set to 0.5 times the Nyquist frequency. In addition, since some studies have shown that the performance difference caused by different random seeds may even be more significant than that caused by different countermeasure system constructions \cite{wang21fa_interspeech}, we repeat the experiment three times for each case of the system using three random seeds, then we calculate the average EER. Table~\ref{tab:example3} suggests that at 0.5 times the Nyquist frequency, the countermeasure systems based on band trimming and bandwidth extension front-ends have similar performance on the ASVspoof2021 LA database. However, the best performance on the ASVspoof2019 LA database is achieved by the baseline front-end system.

\begin{figure}[t]
  \flushleft 
  \includegraphics[width=0.94\linewidth]{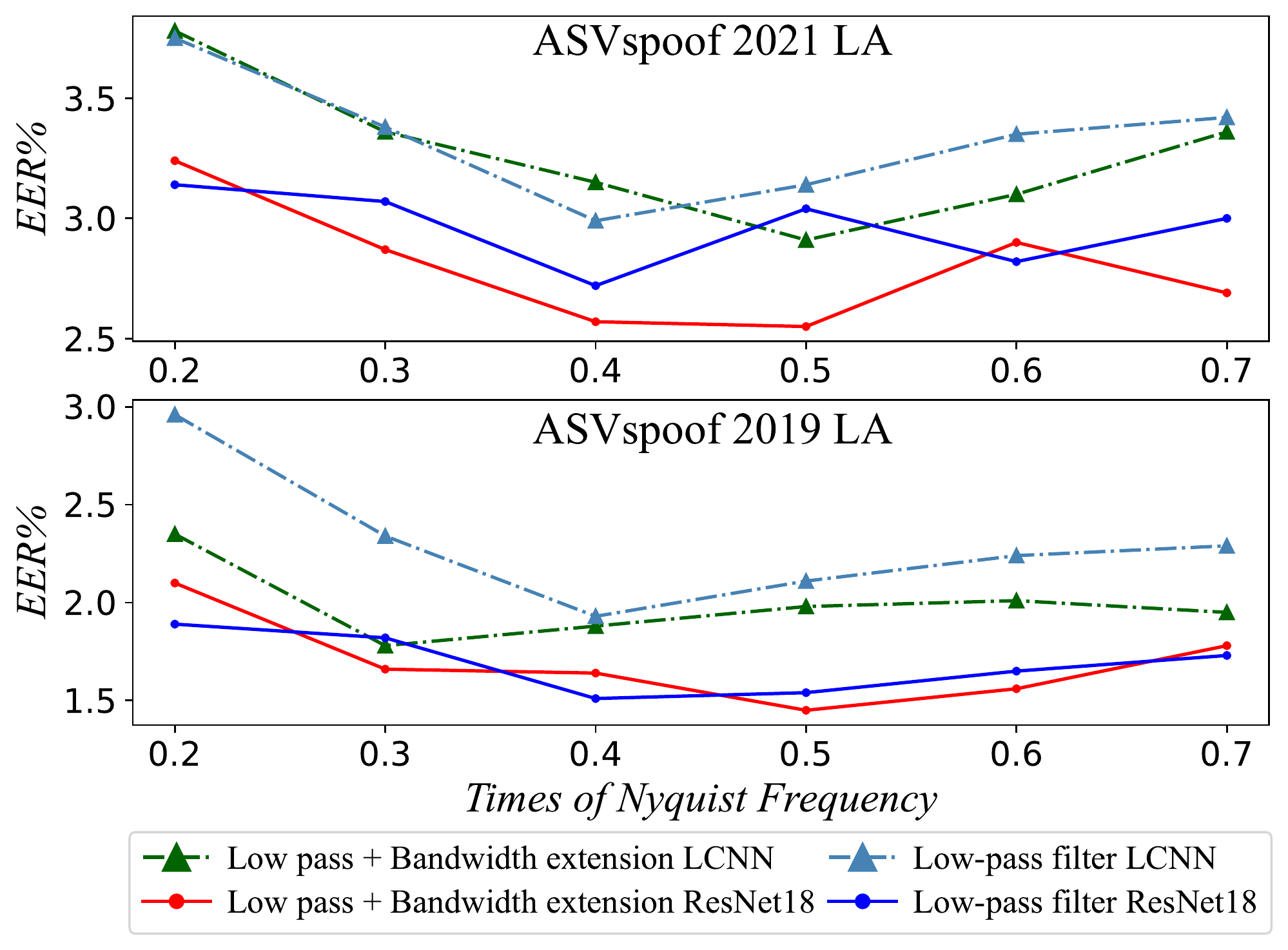}
  \caption{Performance comparison of low-pass filtering and bandwidth extension front-end countermeasure systems. \vspace{-0.6em}}
  \label{fig:result2}
\end{figure}
\begin{table}[t]
\centering
\setlength{\tabcolsep}{4pt} 
\caption{The EER\% of 4 different front-ends and ResNet18 back-end countermeasure systems after VAD operation.}
\label{tab:vad}
\begin{tabular}{llll} 
\toprule
       \textbf{back-ends}&\textbf{front-ends}          & \textbf{21LA}  & \textbf{19LA}  \\ 
\hline
\multirow{4}{*}{ResNet18}    & baseline & 20.17 & 13.06    \\
                             & band trimming     & 19.91  & 18.24 \\ 
                             & low-pass filtering  & 19.08  & 16.64     \\
                             & low pass + bandwidth extension    & 15.23  & 15.08   \\
\bottomrule
\end{tabular}
\vspace{-1.5em}
\end{table}
Under the six cutoff frequencies specified in this experiment, Figure~\ref{fig:result1} provides the results of the countermeasure system based on band trimming and conventional low-pass filter front-end as well as the baselines. It can be found that all countermeasure systems composed of low-frequency subband front-ends outperform the baseline system in the ASVspoof2021 LA database when the cutoff frequency is greater than 0.3 times the Nyquist frequency. As this observation is consistent on two baseline systems, we think it because low-frequency subbands can reduce the variabilities of channel effects and codecs as human ears are less sensitive on high frequency regions which might be more distorted in transmission. The results on the ASVspoof2019 LA database show that low-frequency subbands improve the LCNN backend systems performance significantly more than the ResNet18 system. Combined with the conclusion of Zhang et al. \cite{Zhang}, we suggest that the low pass filtering or down-sampling to narrow band spectrograms is effective when there is relative large channel effects or codec variabilities, however, it might not be useful when there is little high frequency variability on speech. In that case, dropping out high frequency information would result in degraded performance. Figure~\ref{fig:result2} compares the system's performance before and after the bandwidth extension module. It can be found that the bandwidth extension operation on the low pass filtered signal improves performance in most cases. Combining Figure~\ref{fig:result1} and Figure~\ref{fig:result2}, we determine the optimal cutoff frequency for band trimming/filtering/bandwidth extension as 0.5/0.4/0.5 times Nyquist frequency.
\vspace{-0.5em}
\subsection{The effect of VAD operation}

Some studies have shown that since the countermeasure system focuses on silent segments to distinguish spoofed speech from genuine speech, the data after VAD processing will be difficult to classify \cite{Zhang, tomilov21_asvspoof}. We want to further test our proposed methods on the silence removed speech signals after VAD.

We use the librosa.effects.trim function\footnote{http://librosa.org/doc/latest/generated/librosa.effects.trim.html} of the open source toolkit Librosa to implement the VAD function. We set the $top\_db$ parameter to 40, which means that the part of each data with a maximum energy below 40 dB is considered as silence. The performance of each system is shown in Table~\ref{tab:vad} with their own optimal cutoff frequency condition, and it can be found that the EER of all systems increased substantially compared to the ones without VAD. The model with the baseline front-end increase the least on the 2019 database, but the most on the 2021 database; while the EER of systems with the bandwidth extension front-end increased less in both databases. It suggests that bandwidth extension can improve the system performance on speech part of the signal significantly. It also suggest that the original silence signal contains large portion of spoofing artifacts, applying additional signal processing methods on it might distort the silence part artifacts.
\vspace{-0.7em}
\section{Conclusions}
This work validates the countermeasure systems on the ASVspoof2019 LA and the ASVspoof2021 LA databases and shows that low-frequency narrow band can reduce the disturbance caused by channel effects and codec variabilities. In addition, bandwidth extension can significantly reduce the performance degradation after VAD. Moreover, we compared different front-ends and determined the optimal cut-off frequency for those systems. Our future work will focus on exploring the cases that low pass filtering and bandwidth extension are only applied on the speech part of the signal and leave the silence part unchanged.


\bibliographystyle{IEEEtran}

\bibliography{paper}


\end{document}